\documentstyle[12pt]{article}
\begin{document} 

\begin{center}

{\Large {\bf Time and the Mind Body Problem:\\

\medskip

a Quantum Perspective  }}\footnote{Invited conference at the `Psycho-analysis 
and Physics' Meeting. The New School of Social Sciences, New York, December 
1996. In press in \underline{American Imago}.}\\

\medskip

{\bf Jean Schneider\\}
CNRS - Observatoire de Paris, 92195 Meudon, France\\
schneider@obspm.fr
\end{center}

\vspace{2cm}

{\large {\bf 1. The Problem}}

Insofar as we perceive the human body (our own and others) as we do 
any physical 
object, it is 
legitimate to study it with the methods of biophysics, and even of physics.
In this respect, the Mind/Body problem is part of the Mind/Matter problem.

It happens that in today's physics, namely in Quantum Mechanics, the status of 
what an object is is
problematic. My objective here is to show that, as a consequence, the 
point view of the so called cognitive sciences is perhaps to be turned up
side down. I will investigate whether it is pertinent, or not, to see the Mind
as an "emanation from" the material body.

But, first of all, is there anything like "Mind" at all? We live in a technical
epoch where computers  and drugs have the pretension to explain every
mental state in terms of mechanical tropisms. It is thus not useless, for
the sake of the human condition, to recall why a Weltanshauung with no Subject 
cannot explain the totality of everyone's experience. There are several ways 
to conduct such a demonstration, resting for example on the absence of 
demonstrative foundations of the mathematical concepts used by the scientific
theories themselves. I will rather take my argument from the analysis of 
Time. The decisive point is that there is nothing in physics allowing one to
speak of anything like the "now". The only way the express the now lies in
the symbolic dimension of meaning in language. It 
is customary for physicists to 
think of language in terms of communication and exchange of information.
But the dimension of meaning is excluded from such a view. Thus, if one does not
accept the immaterial notion of ``meaning'', one is forced to exclude anything like
the "now" or the present instant.

One could worry about the possibility or the necessity of exploring the 
Mind/Body problem. This necessity is imposed by the empirical fact that we are,
in one way or another, entangled with a body partially described by the
methodology of physics and chemistry (leading for instance to the
pharmacological  care of our body). In Freudian Metapsychology, the Mind/Body
link is ensured by drive. It is the "psychic representative of somatic 
excitations." But Quantum Theory has revolutionized our views of the soma 
which has no more objective properties on its own. It has therefore become 
urgent to 
revisit the Mind/Body problem in the light of quantum concepts.
The interesting outcome of a quantum perspective is that, as we shall see, it 
offers a hypothesis for the way by which a mental representation changes the
/state of the body.

Several authors in the past, for instance Peirce, Heidegger, and Derrida, have
shown that the Mind is intimately involved with time. A rigorous 
investigation of the Mind/Matter problem thus requires some insight into the 
concept
of time. I will therefore make use of some results of recent elaborations of 
this notion, in close connection with Quantum Mechanics.\\

{\large {\bf 2. A Brief Account on Quantum Mechanics and its Problems}}.

Many authors are confused by the formulation of QM in plain language. These 
formulations, if they are taken "\`a la lettre", are often misleading.
For instance, everyone can understand a sentence such as "the uncertainty of
the position of an electron is $\Delta$x". But, taken rigorously, it is
incorrect, as we shall see. It is therefore necessary to state first in a 
correct manner what QM exactly says.\\

{\bf 2.1 The general Framework of QM}.

The conceptual framework of QM can be divided into two parts. It must be
noted that it is almost impossible to explain it correctly in plain language,
unless one makes use of metaphors. These metaphors are most often
fallacious, so that I choose to make use, as simply as possible, of
the language in which physical concepts have been built.
\newpage
\begin{itemize}
\item {\bf Construction and "Natural evolution" of an object}

      The usual textbooks on Quantum Mechanics start with the primitive notion 
      of "system": it is the object under study. But, in fact, an object in 
      Quantum Physics is different from objects of everyday experience:
      it is constructed from sets of experiments. An experiment is the visible
      outcome, expressed in a quantitative way with numbers, of a more or less
      complex setup made of technical devices (\underline{e.g.} a laser source, 
      polarizing filters and photocounters). The behavior of these setups 
      allows one to construct an abstract object like a "photon"
      (or "quantum of light").

      Once the object, or system, is identified, there is attached to it a 
      mathematical object called  "space of 
      representations" $^{\rm (1)}$. To 
      a particular
      configuration of the system is attached an element of this space. This
      element is called "state vector" or "wave function" or simply the "state"
      of the system and is designated by $\psi$.\\

      - Rule {\bf R1}:

      When the system evolves on its own, i.e. is not subject to an 
      observation, $\psi$ changes with time according to some evolution rule,
      called the Schr\"odinger equation, or {\bf S}-evolution.
      The rule {\bf R1} is deterministic and at this level there is no "free 
      will of the atom".

\item {\bf Observation}

      But it is not sufficient to know the mathematical object $\psi$ by
      which the system is represented in the theory. Indeed, $\psi$ is an 
      abstract object, while in the real life of a lab, in their experiments, 
      physicists  only manipulate macroscopic set-ups and numbers, representing
      what is usually called "physical 
      quantities".  By physical quantity, I mean an actual set up in a 
      particular
      configuration, for instance the position of an index on a ruler. 
      Here we have a first level of "free will": the physicist is free to choose
      which "observable" he is going to measure or (equivalently) to observe.
      It is therefore necessary to have rules saying how to compute the 
      numbers representing the result of a given measurement 
      from the knowledge of  $\psi$. 
      There are three such rules:
\newpage
      \begin{itemize}
      \item Rule {\bf R2.1}

      It is a procedure (never mind the details) giving the set 
      of possible outcomes of a measurement. An essential characteristics of 
      the Quantum Theory is that the possible result of 
      a measurement is generally not unique; there is a set of several 
      possibilities 
      (finite or infinite) called the "spectrum" of that quantity for that 
      system. Let a$_1$, a$_2$, etc. be this spectrum (for instance, the 
      different positions of the index of a ruler).

      \item Rule {\bf R2.2}

      The potential, \underline{a priori}, outcome is not unique, but the 
      actual outcome of
      an actual measurement is unique. There is a question here:
      "how is a particular outcome chosen?", "how can we predict it?". The
      answer is that the outcome is random; we can only predict the 
      statistical probability of it. The probability p$_i$ that the outcome
      is a$_i$ can be computed, according to rule R2.2, from
      $\psi$ and a$_i$ $^{\rm (2)}$. If the probability p$_i$ is exactly 1 
      (\underline{i.e.} 100\%) for
      some i$_o$, the outcome is
      certain and can only be a$_{i_o}$ and 
      the system is said to be in the state $\psi$ = (a$_{i_o}$).

      We have now a second level of "free will" in the sense that the outcome
      is not deterministically predictible.

      \item Rule {\bf R2.3}
 
      Just after a measurement has been performed, a new measurement of the 
      same quantity will give the same result a$_i$ with certainty. Thus
      after a measurement with an outcome a$_i$, the state is (a$_i$). 
      In other words, as a result
      of the measurement, the system has {\bf suddenly jumped} from the state
      $\psi$ to the state (a$_i$). This jump, most often called the state 
      vector
      reduction {\bf R} or the wave function collapse, is not deterministic, 
      it has a random result. I call this reduction "evolution of  
      {\bf type R}".

      As we shall see, this rule is the source of all the interpretation
      problems  of the Quantum Theory.
      \end{itemize}
\end{itemize}
      
   There are therefore two types of evolution for a system: the 
{\bf S}-evolution and the {\bf R}-evolution.\\

   Until now there is no problem with this scheme, at least apparently. But
a very serious problem arises when one wants to understand more precisely
what a measurement is. The natural reaction of a physicist is to view it as
a physical interaction between the system and an apparatus. This is an 
\underline{a
priori} reasonable attitude since a physical device is made of atoms. But then
the system under study and the apparatus form together
a metasystem MS which evolves freely according 
to the {\bf S}-evolution, which is deterministic. But the rule {\bf R2.3} 
states that it evolves 
according
to the {\bf R}-evolution whose outcome is only probabilistic. 
Therefore one is led to a contradiction between two
points of view: if a measurement is seen as a "natural" physical process,
its evolution is in contradiction with the rule governing measurements.\\

 Another way to 
express this paradoxical aspect is that, before a measurement, a physical
quantity does not "possess" any definite numerical value (for exemple the 
position of an electron). If one wants to
try a complete physical (in terms of an apparatus-object physical interaction)
description of the measurement, there is no longer any measurement (since there 
is no more an {\bf R}-evolution). This sounds like Zeno's paradox: as soon
as one tries to catch the motion by a rational analysis, there is no more
motion.

What is also 
paradoxical is that the choice between the {\bf S}-evolution and the
{\bf R}-evolution does not  result from a physical process, but from an
arbitrary decision of the physicist who chooses to describe the situation as a
measurement or as a physical object-apparatus interaction. It is because this
arbitrary  decision comes into the game that London and Bauer  proposed 
as early as 1938 that the state vector collapse, or {\bf R}-evolution, is
not a physical process but takes place only in the "observer's consciousness".
I use quotation marks since it is not clear what consciousness is.
It was the introduction of an idealistic wolf into a materialistic sheepfold.

This introduction of an (apparently) idealistic element into physics has 
shocked many physicists. They generally either do no want to discuss these
matters, or  try (a minority of them) to change the axioms of the theory.\\

{\bf 2.2 Some proposed solutions to the measurement problem}.

    The above contradiction between {\bf S}-evolution and {\bf R}-evolution
is, in my opinion, the most crucial problem of
physics. It has triggered many attempts for a solution since the 30's. Let me 
briefly mention, without unnecessary details,  the most elaborate.
\newpage
\begin{enumerate}
\item "Hidden parameters".

     The idea is that the randomness of the {\bf R}-evolution is only apparent.
There are supposed parameters, presently unreachable like the position of the 
atoms of a gas for a 19$^{\rm th}$-century 
physicist, whose evolution governs the
underlying dynamics of the system-apparatus metasystem. It is only because 
these 
parameters have a random statistical distribution, irreproducible from one 
experiment to 
another, like the random, but deteministic, motion of atoms in a gas, that the 
{\bf R}-evolution is, apparently, unpredictible. This proposal has lived for 
several
decades and was a serious alternative to standard QM. In 1966, John Bell
did show that for a very large class of hidden parameters (namely 
\underline{local}
hidden parameters), this kind of theory
leads, in specific cases, to predictions in contradiction for those of QM.
An experiment (on the light emission by atoms) was conducted in 
1983 by Alain Aspect to test whether the new theory or QM was correct. The 
result
was that QM is correct, thus excluding this kind of solution to the 
measurement problem.

\item "No observation" interpretation.

       According to this interpretation, there is no need for an  
{\bf R}-evolution and the rule {\bf R2.3} does not hold. All the possible
outcomes of an experiment (Rule {\bf R2.1}) are simultaneously realized and
there is a set of simultaneous different observers, each one seeing one of the 
outcomes. This interpretation, proposed by Everett, is usually called the many
world interpretation, although it is really a "many observers" interpretation.
It has no internal logical contradiction, but it does not explain a piece of
empirical
evidence: there is always only one \underline{actual}
observer. There is a parallelism 
with time: in physics, there is no privileged instant such as "now" on the 
time line, in contradiction with the empirical evidence of the actual existence
of "now". The reason for these contradictions between the theory and 
empirical evidence is that there is no way to express "actuality" or 
"existence"
in a formal, mathematical way. It is a matter of the foundations of
writing: an equation is a written sign which, as such, is timeless, while
actuality, being an actualization, 
is temporal, in the sense of Heidegger's Urtemporalit\"at.
\newpage
\item Decoherence.

The "decoherence program" has tried to make use of the fact that macroscopic
objects (and thus apparatuses) are assemblies of elementary systems (atoms) 
which
behave incoherently: their wavelike behavior is statistically erased by
destructive interferences of these waves. But this decoherence does not explain 
why in a given experiment there is only one outcome among many 
\underline{a priori}
possibilities. In this respect, it fails to explain the behavior of an 
apparatus as an assembly of elementary quantal constituents (like atoms).

\end{enumerate}

Thus neither of these solutions is satisfactory and we are led to find 
something else in order to understand what the {\bf R}-evolution is.\\

{\large {\bf 3. The \underline{Semiotic Reduction} of the State Vector.}}

Let us go back to the basic formulation of the quantum scheme and to the actual
practice of physicists.

The rule {\bf R 2.1} gives a prescription for the possible outcomes of a 
measurement. We must ask ourselves what, in the real life of a lab, a
 measurement does. More precisely, when, according to which criteria, after
which event, did a measurement really take place? A careful 
phenomenological examination of this process leads to the conclusion that
a measurement is performed when its inscription has taken place. This 
inscription can be either written or oral. In fact it could take place in
any symbolic system. 
In other words, the only "interaction" involved in a measurement is the 
intervention afforded
by a signifier: it is only when a physical system is described in 
words that a measurement can take place. In this view, the classical level 
exists on its own, and the program of decoherence, namely to attain the goal
of understanding the quantum origin of macroscopic appearances is,
by matter of principle, hopeless.
I therefore take as given, as a starting point, that a
measurement is nothing else than its
inscription. J. Von Neumann, one of the founder of QM, did in fact express a
similar view: "Indeed experience only makes statements of this type ``an
observer has made a certain (subjective) observation'', and 
never any like this ``a physical quantity has a certain value'' ". To quote
N. Bohr, "By the word "experiment", we simply refer to a situation where we 
can \underline{tell} $^{\rm (3)}$ others what we have done what we 
have learned."  In fact, the
 word "subjective" is unnecessary here. It is sufficient that an observer 
\underline{states}, or \underline{declares} what he has observed. In fact in 
this statement, in this \underline{act} of statement I should say, the observer
as a subject disappears behind his statement, and the latter thus
acquires an 
objective, or more precisely intersubjective, status. Thus, in real practice, 
an observation, or a measurement, is identical to its declaration. It is
intersubjective in the sense that a declaration is always shared by the 
interlocutors. The communication scheme, according to which 1/ I first have in
mind something I want to communicate, and 
2/ in a second time what was in my mind 
is after the communication transfered in the mind of my audience,
is in my opinion wrong. The real situation is more atemporal. Time in the mental
world is not adequately represented by the real variable \underline{t}, 
it involves 
"afterwardness", having thus a non linear character (which can be formalized 
rigorously [Schneider 1994 and references therein]) 
and an instant is not a point. In the framework 
of this mental time, I get an idea
in my mind only afterwards, when my interlocutor has received it. Finally,
phenomenologically speaking, an observation, as a declaration, only exists in
the impersonal universe of discourse. 

One can make the following analogy with linguistics or with mathematics.
These sciences study statements (of natural language or of mathematics) for 
themselves, and never either for the 
psychological motivations or detailed physical
mechanisms (acoustical for instance) underlying their production.

The question of understanding how 
individual subjects are articulated, or linked, to the universe of discourse 
is another 
problem which I will not discuss here and which requires more elaborated 
notions such as various identifications, incorporation and introjection, dual 
unity (Ferenczi, Melanie Klein, Nicolas Abraham, Aulagnier).\\

To summarize, I adopt the hypothesis that

\begin{center}
{\bf The R-evolution is a Speech Act}
\end{center}
in the sense of Austin, that is, the act of production of a symbol,  
and not a physical (i.e.
independent from an observer) phenomenon since
physical phenomena strictly depend on {\bf S}-evolutions. In the present case, 
this symbol is a 
mathematical writing, or, to be more precise, the reading of a written formula. 
Like any writing, it is not a static trail. As a verb, a 
writing is a gesture, the gesture of a production of a symbol. A symbol 
is homogeneous to its own production, it \underline{is} its own production.
This production takes place in the transcendental time of the emergence of an 
instant. It can be demonstrated (Schneider 1994)
that a transcendental instant necessarily has a certain chronological duration
$\Delta$T, in terms of our watches. This discretization of transcendental time
was already discussed, in different terms, by several authors. Heidegger, for 
instance, names it "spannung" and "entstreckung" of the present.  I do not 
say "in terms of physical time". 
Indeed, 
it is misleading to identify the mathematical variable \underline{t} 
used by physicists
with time in the phenomenological sense of this word. The \underline{t}
variable, at least
for values very shorter than $\Delta$T ("very short times"), is never directly 
apprehended as time, 
but as a some number deduced from seizable
quantities such as a length \underline{L} which
is converted afterwards into \underline{t} through relationships such as 
\underline{t} = \underline{L}/\underline{v} (where 
\underline{v} is a velocity).

This  theory of a transcendental time has no way to predict 
\underline{a priori} 
the value of the quantum $\Delta$T.
It is an empirical fact, in reality unexplained (even by psycho-physiology),
that $\Delta$T is approximately 0.1 sec.

The combination of the existence of the "lapse" $\Delta$T attached to any
signifier and of
the hypothesis that a measurement is the production of a signifier (belonging 
to the world of physical quantities) leads to a definite prediction: a 
measurement in Quantum Physics cannot be shorter than $\Delta$T.\\

After a reduction {\bf R} has taken place, it is not objectively stalled for
ever.
If a second observer asks the first one what is the result of his 
measurement, the answer of the first observer produces a new statement and 
thus a second reduction {\bf R}. This evanescence  of the reductions {\bf R}
is perhaps similar to the evanescence of the present: once an instant has been
 "presentified", it is immediately superseded by another.\\

This model of a semiotic state vector reduction provides moreover an 
explanation of
an embarrassing fact, namely that it is not compatible with relativity 
theory (because it is instantaneous everywhere or, equivalently, taking
place with an infinite speed). This 
incompatibily is well explained by the semiotic state vector reduction:
the semiotic state vector reduction being a phenomenon of discourse, it takes
places in the universe of interlocution which provides a privileged, unique,
 frame of
reference. Such a privileged frame does not exist in the physical world, 
according to the theory of relativity, where all frames are equivalent.

\newpage
{\large {\bf 4. The Mind/Body Problem: Toward a Quantum Psychology}}

{\bf (A model for an action of mental representations on the physical body)}.\\

In the past, a few authors, physicists and others, have elaborated some 
thinking on the
relationship between QM and the Mind/Matter problem. Some physicists have
built quantum models of the Mind/Body relation. A useful partial account
can be found in Stapp (1993 and 1997). For instance, Costa de
Beauregard (1966) has proposed that mental representations can change the 
probabilities in rule {\bf R 2.2}. More recently, Hameroff and Penrose have
tried to
1/ make an objective description of the {\bf R}-evolution and 
2/ propose that this
{\bf R}-evolution \underline{is} consciousness. In the humanities, some
philosophers, psychologists or psychoanalysts and sociologists have used the 
QM analogy as a model of the subject-object interaction or relationship
(transference in the case of a psychoanalytical viewpoint).
(See for instance S. Viderman, J. Laplanche, S. Felman). This
analogy should be investigated carefully 
since the status of this interaction is slightly different in QM and 
in the human world.\\

The idea of a semiotic state vector reduction is essentially that the reduction
{\bf R} is the production of a signifier. This production creates the result of
the measurement (instead of recording passively a matter of fact like in 
Classical Physics) and changes 
the state of the system, according to the rule {\bf R 2.3}. Since we have up 
to now dealt with physical measurements, the signifiers which are produced 
belong to the universe of physical concepts.

But the symbolic register has more dimensions than just conceptual ones. It
is much larger and includes signifiers (being possibly unconscious) having 
esthetical, affective, ethical,
etc. values. I make this statement in parallel with the notion of drive
in Freudian Metapsychology
where, for instance, I understand the destiny of the epistemophilic drive 
(as a sublimation) 
as the production of a signifier having a conceptual value. Like the 
drive, any signifier takes its source from our soma. But what is the
soma in QM?. Before QM, it was an objective object. In QM, we have two
heterogeneous notions, as we have seen in part {\bf 2}:\\

- the object under study (described by its state vector)

- the properties or attributes given to the object by our perception of it.\\
\bigskip
Which one is our body? The state of the object, or its properties, or both?
I take the view that:\\

- as a physical object, the body is a system in the Quantum Mechanical sense
   (an assembly of atoms) with no qualities\\

- as a phenomenal object, it is a system plus its attributes given to it
by our relationship to it.\\
\bigskip\\
If our relationship is a measurement, we get physical attributes; if it
is an affective relationship, we get symptoms.

For the 
physicist, the properties are descriptions of the system in terms of
physical concepts (or signifiers). But for the
phenomenological approach of each one of us, with our phantasmatic (sometimes 
unconscious) representations of our body, it has appearances 
("properties") such as: beauty, 
erotization of bodily zones, partial identifications, etc., for which physical 
concepts 
are inadequate.\\

Let me give some specific examples.

\begin{enumerate}
\item {\bf "Voluntary" action}

      When I decide to raise my hand, there is a global representation 
of this action in my mind. This representation does not have the form of a
detailed physical description such as ``the muscular fibers of this arm will
contract at such or such strength, etc.'' There are global signifiers
"motion of my arm from position A to position B". In the present model,
we have:

- the arm as a quantum system with no particular property

- each
slight motion is then the result of the "state vector reduction" associated 
with the production of signifiers such as the above. 

The accumulation of these 
slight motions gives the global motion of the arm.
\item {\bf Blushing of the face}

       Let us consider a signifier with an affective value. As an 
(affective) representation, and thus production of an (affective) signifier, 
it can, as an {\bf R}-evolution applied to the system "blood circuits of the 
face", change the apparent vascularization of the face, the latter appearing 
thus as blushing. 

       A similar model can be invoked for hysterical somatisation or
for psychosomatic  phenomena (such as some allergies) when the 
signifier is unconscious or preconscious. It is irrelevant here to point out
the disctinction between hysteria and psychosomatic affections. It is true that
in the case of hysteria there is no visible physiological affection, but at 
the end, the entire body behaves differently as when there is no symptom.\\

\end{enumerate}
How could such a model be tested? One should be able to find two different
signifiers S$_1$ and S$_2$ which are incompatible in the quantum sense, i.e. 
such that after an outcome for S$_1$ the outcome for S$_2$ is unpredictible.
A question arises here. Since psychological signifiers always refer to our body
which is a macroscopic system, can they be constructed as "collective 
variables"?

A noticeable consequence is that there is no more "psychophysical parallelism". 
Indeed, the mental world is represented by the measurement operators (and more
exactly their proper values), while the physical body is represented by
the state vector. There is no parallelism between them. \\
\bigskip

The new hypotheses outlined above can serve as first steps for future 
developments:

\begin{itemize}
\item read the classical important texts in metapsychology on the Mind/Body
relation through  the lens of Quantum Physics.
\item investigate the "collective" observables (such as "morphology" and other 
qualitative features) given to the system.
\newpage
\item investigate whether Quantum Physics can be extended to observables whose 
 result are not 
expressed in numerical terms. This is important for the foundation of a truly
quantum psychology.\\

\end{itemize}

\bigskip
I am grateful to Alan Bass for his help for the english style.\\

\bigskip

{\bf Notes:}

(1) For experts, it is a Hilbert Space.

(2) Namely p$_i$ = F($\psi$, a$_i$)       
      where F is, never mind the details, a simple function having 
      some given standard form.

(3) Underlined by me.\\

\bigskip

{\bf References.}\\

Abraham, Nicolas and Torok, Maria. 1978. L'\'ecorce et le noyau. Paris: 
Aubier-Flammarion. The Shell and the Kernel. Translated by Nicholas Rand. 
Chicago: Chicago University Press.

Aulagnier, Piera. 1984. L'apprenti-historien et le ma\^{\i}tre-sorcier. Paris:
Presses Universitaires de France.

Austin, John. 1962. How to do things with words. Oxford: Oxford University
Press.

Bohr, Niels. 1949. Discussions with Einstein on epistemological problems in
atomic physics. in \underline{Einstein Philosopher ans Scientist.} 
Arthur Schilpp Ed. Evanston: The  Library of Living Philosophers.

Costa de Beauregard, Olivier. 1966. Le second principe de la science du temps 
Paris: Seuil.

Derrida, Jacques. 1970. Marges de la philosophie. Paris: Minuit.
Margins of philosophy. Translated by Alan Bass.  Chicago : University of 
Chicago Press.

Felman, Shoshana. 1978. Le scandale du corps parlant. Paris: Seuil.

Ferenczi, Sandor. 1952. First contributions in Psycho-analysis. London: Hogart 
Press.

Hameroff, Stuart. 1995. Quantum Coherence in Micro-tubules: a Neural Basis for 
Emergent Consciousness, in \underline{Journal of Consciousness Studies}. 1:91.

Heidegger, Martin. 1972. Zur Sache des Denkens. Heidelberg: Niemeyer.
On Time and Being. Translated by 
Joan Stambaugh. New York: Harper and Row.

Klein, Melanie; Heimann, Paula; Isaacs, J. and Riviere, Joan. 1952.
Developments in Psycho-analysis. London: Hogart Press.

Laplanche, Jean. 1970. Vie et mort en psychanalyse. Paris: Flammarion.
Life and death in psychoanalysis. Translated by Jeffrey Mehlman.
Baltimore: Johns Hopkins University Press.

London, Fritz and Bauer, Edmond. 1938. La th\'eorie de l'observation en 
m\'ecanique 
quantique. Paris: Hermann. Translated in \underline{Quantum Theory of}
\underline{Measurement.}
John Archibald Wheeler and Wojciek Zurek, Eds. Princeton: Princeton University 
Press.

Peirce, Charles Sanders. 1982 Writings of Charles Sanders Peirce. Volume 3. 
Edited by  Max H. Fisch. Bloomington: Indiana University Press.

Penrose, Roger. 1989. The Emperor's New mind. Oxford: Oxford University Press.

Schneider, Jean. 1994. The Now, Relativity Theory and quantum Mechanics.
in \underline{Time, Now and Quantum Mechanics.} 
Michel Bitbol and Eva
Ruhnau Eds. Gif-sur-Yvette: \'Editions Fronti\`eres.

Stapp, Henry. 1993. Mind, Matter and Quantum Mechanics. Berlin: Springer.

Stapp, Henry. 1997. Why Classical Mechanics Cannot Naturally Accomodate
Consciousness But Quantum Mechanics Can. to appear in \underline{Psyche.}

Viderman, Serge. 1970. La construction de l'espace analytique. Paris: Deno\"el.

Von Neumann, John. 1932. Mathematische Grundlagen der Quantenmechanik.
Berlin: Springer. Mathematical Foundations of Quantum Mechanics. 
translated by Robert Bayer. Princeton: Princeton
University Press.\\

\bigskip

Jean Schneider

Observatoire de Paris - CNRS, 92195 Meudon, France

http://www.obspm.fr/departement/darc/schneider/qm.html

\newpage

\begin{center}
{\Large {\bf Appendix}}\\
\end{center}

\medskip

\begin{center}
{\large {\bf Indeterminism, Causality, Complementarity.}}\\
\end{center}

\bigskip

Quantum Mechanics has given rise to many comments in the field of humanities,
most of them misleading. Non-causality, complementarity, and wave particle 
dualism are often refered to in unclear terms. These vague notions 
are not really used in the technical works of physicists (even if they use
them for the layman). What is it really about?\\

\underline{Indeterminism} only means that, before a measurement, one can 
generally not predict its outcome. This word refers only to the measurement 
and not to the behavior of quantum objects on their own, which is entirely
deteministic (being governed by the {\bf S}-evolution).\\

\underline{Complementarity} has, somehow, a precise mathematical 
formulation, namely the Heisenberg uncertainty relation $\Delta$A.$\Delta$B
$\geq$ h. But it is necessary to interpret the latter correctly.

This relation bears exclusively on "ensembles", identical copies of a given
system, not on individual systems. $\Delta$A (or $\Delta$B) is not to be
interpreted as the "uncertainty" of the knowlege of the quantity A (or B).
The word uncertainty does not have in Physics the same meaning as in 
ordinary
language; in ordinary language it means that an entity has a given value but 
that we do not know what that value is. In Quantum Physics, a physical quantity 
{\bf A}, as a concept, does not possess a given value A on its own; only the 
\underline{result} of the measurement of that quantity exists. 
$\Delta$A is then only
 the statistical dispersion, around a mean value, of all the different 
measurements when repeated on 
many copies of the same system (in the same state).
 The smaller $\Delta$A is, the better the precision on the mean value of
{\bf A}. 

"Complementarity" then means nothing else than the statistical dispersions
$\Delta$A and $\Delta$B cannot both 
be infinitely small for the same state of the 
system under consideration. The
expression "the value of {\bf A}" has no ontological meaning and the
statement "{\bf A} has the value A" has no meaning in QM. The only
meaningful sentence is "the measurement of {\bf A} has given, or is predicted 
to give, a result A". This emphasis on measurement rests on the same grounds
as those of Relativity Theory which forbids one to refer ontologically 
to the position or 
to the instant of an event: only their measurements have a meaning.\\

The \underline{wave-particle dualism} rest on a fact discovered by Louis de
Broglie in 1923 - 1924: if
a system is in a state $\psi$ such that measurement of its impulse has an 
exactly predictible outcome P, then the mathematical expression of $\psi$ is 
a wave with a wavelength $\lambda$ = h/P (This is not a new postulate,
it is a consequence of the rule {\bf R2.1}; h is Planck's constant). In this
case, $\psi$ has a spatially infinite extension and the measurement of the 
position can give, with equal probability, a result anywhere in space. In that 
sense, \underline{and only in that sense}, the system has no assignable 
position. 
If the precision on P vanishes, so does, as a consequence of $\lambda$ = h/P,
the precision on $\lambda$ and the wave-like nature (or, more exactly,
appearance) of $\psi$ fades out.
As a particular case of $\Delta$A.$\Delta$B $\geq$ h, one has here
$\Delta$x.$\Delta$P $\geq$ h (where x is the position of the object under 
study). Thus, as the imprecision $\Delta$P on P (to be
more exact the statistical dispersion on repeated measurements)
increases, the imprecision $\Delta$x on x decreases. At the limit
where $\Delta$P becomes infinitely large, the 
prediction of the results on position measurements becomes 
infinitely precise, a way to see the
object as "punctual" or corpuscular. The "wave-particle duality" is thus not 
ontological, but is the fact that, on a given system, the predictions on the
position \underline{measurements} and the impulse \underline{measurements} 
cannot be both 
infinitely precise.\\

\underline{Non-causality}. The "natural" evolution of a system (without 
measurement) is described by the Schr\"odinger equation and is absolutely
deterministic. The non-causality, or the impossibility of making a prediction,
bears only on the outcome of measurements, and is possible because measurements 
do not follow the Schr\"odinger equation.\\

\underline{Observer-system interaction}. It is often said that complementarity,
or indeterminism, comes from the uncontrollable influence of the observer on 
the observed system. This is not completely wrong, but, at the same time, also 
not completely correct. What is the "influence" of the observer on the system?
It is natural (especially for a physicist) to view this influence as a material
interaction between the system S and the observer (or its apparatus)
O. But then 
this interaction would be described by a Schr\"odinger equation bearing on the
meta-system  S + O. Since the Schr\"odinger evolution of 
S + O is deterministic, the invoked "influence" is in turn
deterministic and thus not "uncontrollable": there is no possibility of
explaining by this way the uncertainty $\Delta$A on the measurement of {\bf A}.
The observer's influence on the system resides only in the, unpredictible,
wave-packet collapse (which, in the context of the present study, is
of semiotic
nature). For the semiotic state vector reduction, there is only an interaction 
between symbols and the system represented by $\psi$.
This interaction can tentatively serve as a prototype for the Mind/Body
relation.
\end{document}